\documentclass[a4paper]{article}
\pdfoutput=1
\usepackage{graphicx}
\usepackage{amsmath}
\def\e#1{\ensuremath{_{\text{#1}}}}

\begin{document}

\begin{center}
\LARGE Mechanism and kinetics of hydrated electron diffusion \\[5mm]
\normalsize Kafui A. Tay, Fran\c{c}ois-Xavier Coudert$^{*}$, and Anne Boutin \\
\footnotesize Laboratoire de Chimie Physique, Universit\'e de Paris-Sud, 91045 Orsay Cedex, France. \\
$^{*}$Present address: Davy Faraday Research Laboratory, UCL, Gower Street, London WC1E~6BT, United Kingdom.
\end{center}

\begin{abstract}
\noindent
Molecular dynamics simulations are used to study the mechanism and kinetics of hydrated electron diffusion. The electron centre of mass is found to exhibit Brownian-like behaviour with a diffusion coefficient considerably greater than that of the solvent. As previously postulated by both experimental and theoretical works the instantaneous response of the electron to the librational motions of surrounding water molecules constitutes the principal mode of motion. The diffusive mechanism can be understood within the traditional framework of transfer diffusion processes, where the diffusive step is akin to the exchange of an extra-molecular electron between neighbouring water molecules. This is a second-order process with a computed rate constant of 5.0~ps$^{-1}$ at 298~K. In agreement with experiment the electron diffusion exhibits Arrhenius behaviour over the temperature range 298~K--400~K. We compute an activation energy of 8.9~kJ mol$^{-1}$. Through analysis of Arrhenius plo
 ts and the application of a simple random walk model it is demonstrated that the computed rate constant for exchange of an excess electron is indeed the phenomenological rate constant associated with the diffusive process.
\end{abstract}

\newpage

\normalsize
\section{Introduction}

\noindent
The hydrated electron has long been a topic of fundamental interest~\cite{hart,jortner,gould,kevan}. Since its original identification~\cite{hart62} it has become a focus of study particularly pertinent to radiolytic and electron initiated processes~\cite{pikaev}. Ionisation of a polar solvent, by radiolysis or photolysis, leads to the production of a solvated electron. This transient species may localize and thermalize in the solvent, eventually being captured in a potential energy well formed by the solvating molecules. Whilst being all the time solvated, fluctuations in the solvent coordinate enable diffusion through the solvent medium. Upon encountering an acceptor species of sufficient electron affinity the electron may be incorporated into one of its orbitals. In such cases the rate of electron transfer or electron-initiated processes is inextricably linked to the diffusive dynamics of the solvated electron~\cite{bartels95}.

The hydrated electron, identified by its characteristic optical absorption spectrum~\cite{hart62,freeman79}, has been closely studied in ambient conditions by pump-probe spectroscopy~\cite{antonetti87,eisenthal90,barbara93,barbara98a,barbara98b,wiersma99,laubereau00,hertwig02,nilsson07}. These works reveal a sophisticated picture of the early-time, ultra-fast dynamics of solvation from localisation to equilibration and have been well complemented by computer simulations~\cite{rossky86,berne87,landman91}. The diffusive dynamics following equilibration have also been a focus of interest. Conductometric measurements of electron diffusion in water reveal both an unusually high mobility (several times greater than that of classical ions) and Arrhenius-like diffusion~\cite{bartels95,bartels92}. Despite numerous computer simulations the fundamental physics of hydrated electron diffusion still remains poorly understood~\cite{nitzan89,rossky89,murphrey92}. Grotthus type mechanisms, ad
 iabatic long-range hopping, and tunnelling effects have all been discounted. Instead, a polaron-like picture of the dynamics has emerged in which the high mobility has been attributed to the instantaneous adiabatic response of the electron to fast solvent motions~\cite{nitzan89,rossky89,lakhno07,park01}.

In this work we demonstrate that the mechanism and kinetics of the diffusion process are well described by a transfer diffusion model~\cite{ruff71,dahms68}. We consider a transfer process involving the exchange of an extra-molecular electron between identical solvent molecules $A_{1}$ and $A_{2}$, 
\begin{equation}
\label{equation1}
[A_{1},e^{-}\e{aq}] + A_{2} \rightleftharpoons  A_{1} + [A_{2},e^{-}\e{aq}].
\end{equation}
For a system populated entirely of $A_{1}$ and $A_{2}$ migration of the excess charge occurs either by the above charge-transfer process and/or by the conventional diffusion of $[A_{1},e^{-}\e{aq}]$ or $[A_{2},e^{-}\e{aq}]$. If the timescales associated with the diffusion of $A$ and the electron-transfer process are similar in magnitude both processes may contribute significantly to the diffusion of charge. For example, voltammetric studies of electron conduction in inorganic solutions indicate that the charge-transfer process provides a significant contribution to the overall migration of charge~\cite{norton92}.

We employ an adiabatic quantum-classical molecular dynamics method to investigate hydrated electron diffusion over the temperature range 298~K to 400~K. In common with other works~\cite{nitzan89,rossky89,murphrey92} the system is evolved within the Born-Oppenheimer approximation with the solvent molecules treated classically. The electron, confined to the ground state, is represented by a wavefunction that corresponds to the instantaneous nuclear configuration. Our methodology has been previously employed for the hydrated electron at different thermodynamic states~\cite{boutin03a,boutin05}, in confined media~\cite{boutin06a}, and in electron-cation pairs~\cite{boutin06b,boutin04b,boutin03b,boutin07}.

\section{Method}
Mixed quantum-classical molecular dynamics is used to simulate the excess electron in bulk water~\cite{boutin03a,borgis95}. The problem of the hydrated electron is reduced to a localised quantum mechanical problem within a classically evolving system~\cite{webster91}. Water-water interactions are treated classically, using the SPC model~\cite{spc}, with the solvent coordinate, $\mathbf{S}$, evolved according to Newton's equation of motion. The electron-solvent interaction potential of a given configuration is termed $\hat{V}\e{int}(\mathbf{r},\mathbf{S})$, where $\mathbf{r}$ denotes the electronic coordinates. The energies ($E_{n}(\mathbf{S})$) and wavefunctions ($\psi_{n}(\mathbf{r},\mathbf{S})$) of the excess electron may be computed by solving the time-independent Schr\"odinger equation,
\begin{equation}
\label{equationa}
\left[\hat{T}\e{e}+\hat{V}\e{int}(\mathbf{r},\mathbf{S})\right]\psi_{n}(\mathbf{r},\mathbf{S})=E_{n}(\mathbf{S})\psi_{n}(\mathbf{r},\mathbf{S}) ,
\end{equation}
where $\hat{T}\e{e}$ is the electronic kinetic energy operator. The electron-water interaction is described by the pseudopotential of Turi and Borgis~\cite{borgis02} which has been designed specifically for liquid simulations. The principal advantages of this pseudopotential are its simple analytical form, and its good reproduction of both the equilibrium ground state energy and experimentally measured quantities (e.g. optical absorption spectrum).

The Born-Oppenheimer approximation, from which we assume the separation of timescales between the electron and the classical subsystem, allows for an adiabatic simulation of the excess electron. Thus the electron wavefunction is confined to the ground electronic state corresponding to the instantaneous configuration of the nuclear (solvent) coordinates. The quantum mechanical contribution to the forces acting upon the solvent from the excess electron are computed according to the Hellmann-Feynman theorem~\cite{feynman,hellmann}: 
\begin{equation}
\label{equationb}
\mathbf{F}^{(n)}_{S}=-\frac{\partial}{\partial S}E_{n}(\mathbf{S})=-\int^{+\infty}_{-\infty}d\mathbf{r}\quad \left|\psi_{n}(\mathbf{r},\mathbf{S})\right|^{2}\frac{\partial}{\partial S}\hat{V}_{int}(\mathbf{r},\mathbf{S}) .
\end{equation}
We compute only the contributions arising from the ground state ($n=0$) of the excess electron. We note that the energetic difference between the ground and the first excited state of the excess electron is sufficiently large so as to render any thermal excitation negligible. The adiabatic mixed quantum-classical methodology outlined above is sufficiently inexpensive so as to allow the relatively long simulation times ($\sim$0.5 nanoseconds) necessary for investigating the diffusive dynamics. Indeed one electron models, of the kind presented here, have been shown to provide a good approximation to multi-electron approaches, despite omitting the spreading of the excess electron density into the anti-bonding orbitals of hydrating water molecules~\cite{shkrob07}. For a more detailed presentation of our methodology the reader is referred to the relevant literature~\cite{boutin03a,borgis95,borgis02}.

The excess electron was simulated in a bulk system (density 1.00~g/cm$^{3}$) of 300 water molecules at 298~K. Simulations were performed within the canonical ensemble using the Nos\'{e}-Hoover thermostat~\cite{hoover}. Periodic boundary conditions were employed and long-range forces treated using the Ewald summation technique~\cite{tildesley}. The Gear predictor-corrector algorithm was used to integrate the translational and rotational equations of motion~\cite{tildesley,gear}. After equilibration the system was simulated for 500 picoseconds, with a timestep of 0.5 femtoseconds. Configurations were sampled every 5 femtoseconds. This process was repeated at progressively higher temperatures: 325, 350, 375, and 400~K. All simulations were performed at the same density of 1.00~g/cm$^{3}$. Additional simulations were performed on bulk water systems without the excess electron.

\section{Results and Discussion}
\begin{figure}[htb]
\includegraphics[width=8cm]{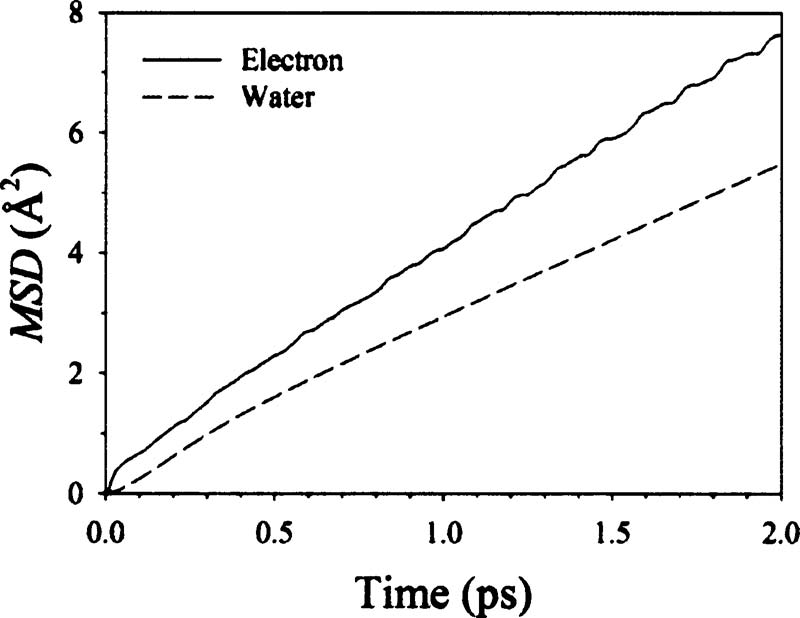}
\caption{Mean-square displacement of hydrated electron centre-of-mass compared with that of bulk SPC water.}
\label{fig:figure1}
\end{figure}

As previously demonstrated~\cite{boutin03a,borgis02}, the hydrated electron is found to be a well-localized entity in agreement with experiment~\cite{kevan81,jortner73} and other simulation works~\cite{nitzan89,rossky89,murphrey92,borgis95,boero07}. As such the electron position may be taken as its centre of mass, $r\e{e} = \left<\psi\left|r\right|\psi\right>$. Though the electron wavefunction is not propagated in time we may examine the electron's dynamical behaviour semi-classically through its centre of mass. Figure~\ref{fig:figure1} compares the mean-squared displacement, $\left\langle \Delta r\e{e}(t)^{2}\right\rangle$, of the hydrated electron with bulk SPC water at 298~K. The proportionality between $\left\langle \Delta r\e{e}(t)^{2}\right\rangle$ and $t$ after 1~ps, indicates that the electron undergoes Brownian-like diffusion. Assuming Brownian motion, the related diffusion coefficient is given by the Einstein relation, 
\begin{equation}
\label{equation2}
\frac{d \left\langle \Delta r\e{e}(t)^{2} \right\rangle}{dt} = 6Dt,
\end{equation}
where $\Delta r\e{e}(t)$ is the total displacement between $t = 0$ and $t$, and $D$ is the respective coefficient of diffusion. The electron diffusion, found here to be 0.60~\AA$^{2}$ps$^{-1}$, is approximately 1.6 times faster than the SPC solvent, 0.38~\AA$^{2}$ps$^{-1}$. In contrast to early simulation work the electron diffusion is clearly greater than the solvent~\cite{rossky89,murphrey92}. Using a similar methodology with the rigid, polarisable TIP4P model for water~\cite{jorgensen83}, Staib and Borgis have estimated the electron diffusion at 0.4~\AA$^{2}$ps$^{-1}$~\cite{borgis95}. Both here and in the work of Staib and Borgis fast diffusion occurs without the inclusion of tunneling effects. Long-range hopping effects are also absent. Conductometric experiments estimate the hydrated electron diffusion at 0.49~\AA$^{2}$~\cite{bartels95,bartels92} approximately 2.1 times faster than water self-diffusion, 0.23~\AA$^{2}$ps$^{-1}$~\cite{kauzmann}. 

\begin{figure}[htb]
\includegraphics[width=8cm]{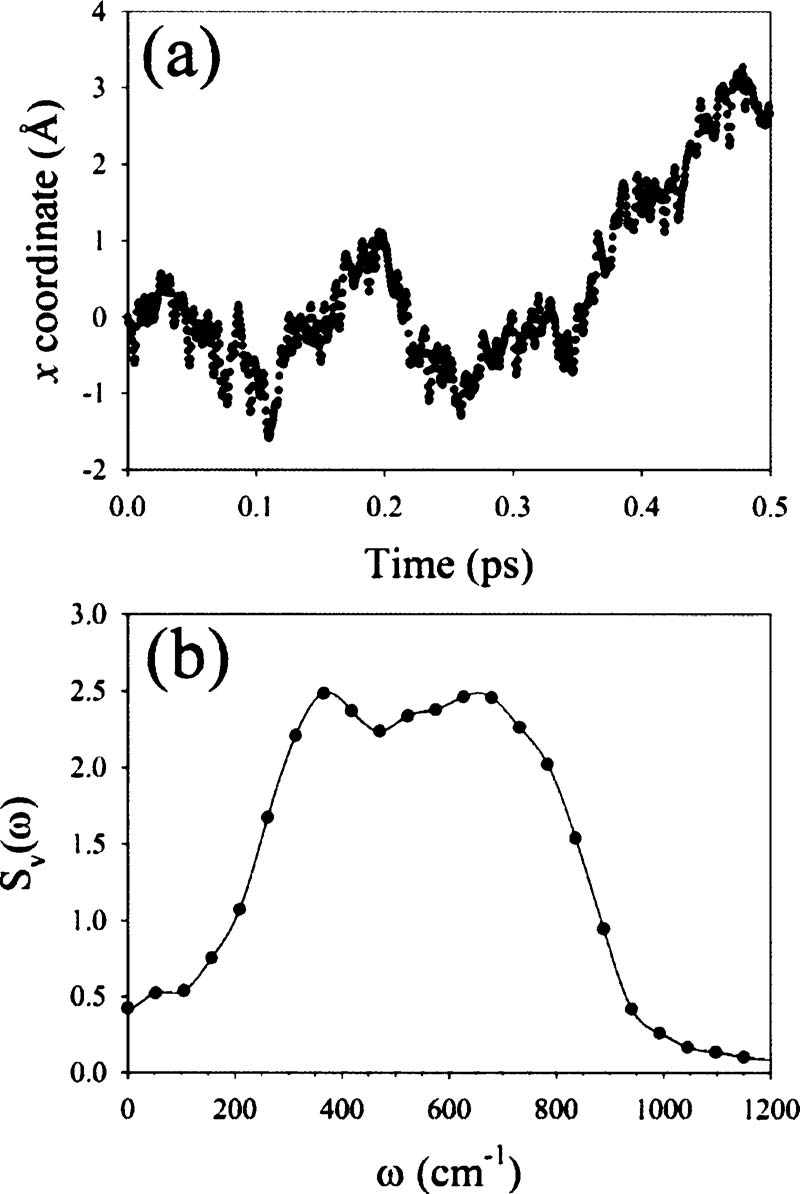}
\caption{Top, $x$-coordinate of the electron centre of mass as a function of time. Bottom, power spectrum of the electron velocity autocorrelation calculated according to equations~\ref{equation3} and~\ref{equation4}. The curve is obtained by cublic spline interpolation.}
\label{fig:figure2}
\end{figure}

A direct examination of the electron position reveals not only the long-time diffusive dynamics but also the shorter-time, oscillatory motions (Figure~\ref{fig:figure2}a). The instantaneous response of the electron to the solvent coordinate correlates the electron and solvent dynamics. As such we may assume that the timescales of the electron dynamics mirror the individual and collective modes of the surrounding water molecules. We determine the frequencies of the electron motion from the spectral density, $S_{v}(\omega)$, of the electron centre of mass velocity auto-correlation function $C_{v}(t)$,
\begin{equation}
\label{equation3}
S_{v}(\omega) = \int^{\infty}_{0} C_{v}(t) \cos(\omega t) ~dt,
\end{equation}
\begin{equation}
\label{equation4}
C_{v}(t) = \frac{\left\langle \textbf{v}(0)\cdot \textbf{v}(t) \right\rangle}{\left\langle \textbf{v}(0) \cdot \textbf{v}(0) \right\rangle}, \qquad
\textbf{v}(t) = \frac{\textbf{r$_e$}(t+\Delta t/2) - \textbf{r$_e$}(t-\Delta t/2)}{\Delta t}.
\end{equation}
The electron velocity, $\textbf{v}(t)$, is computed from the particle displacement between sample intervals. $S_{v}(\omega)$ is shown in Figure~\ref{fig:figure2}b for the hydrated electron at 298~K. The peak frequencies 376~cm$^{-1}$ and 653~cm$^{-1}$ (extracted through cubic interpolation) lie within the region characterised by intermolecular vibrations in bulk water. Low frequency Raman spectra of the hydrated electron exhibits two main peaks at around 450~cm$^{-1}$ and 700~cm$^{-1}$~\cite{tauber03}.

\begin{figure}[htb]
\includegraphics[width=8cm]{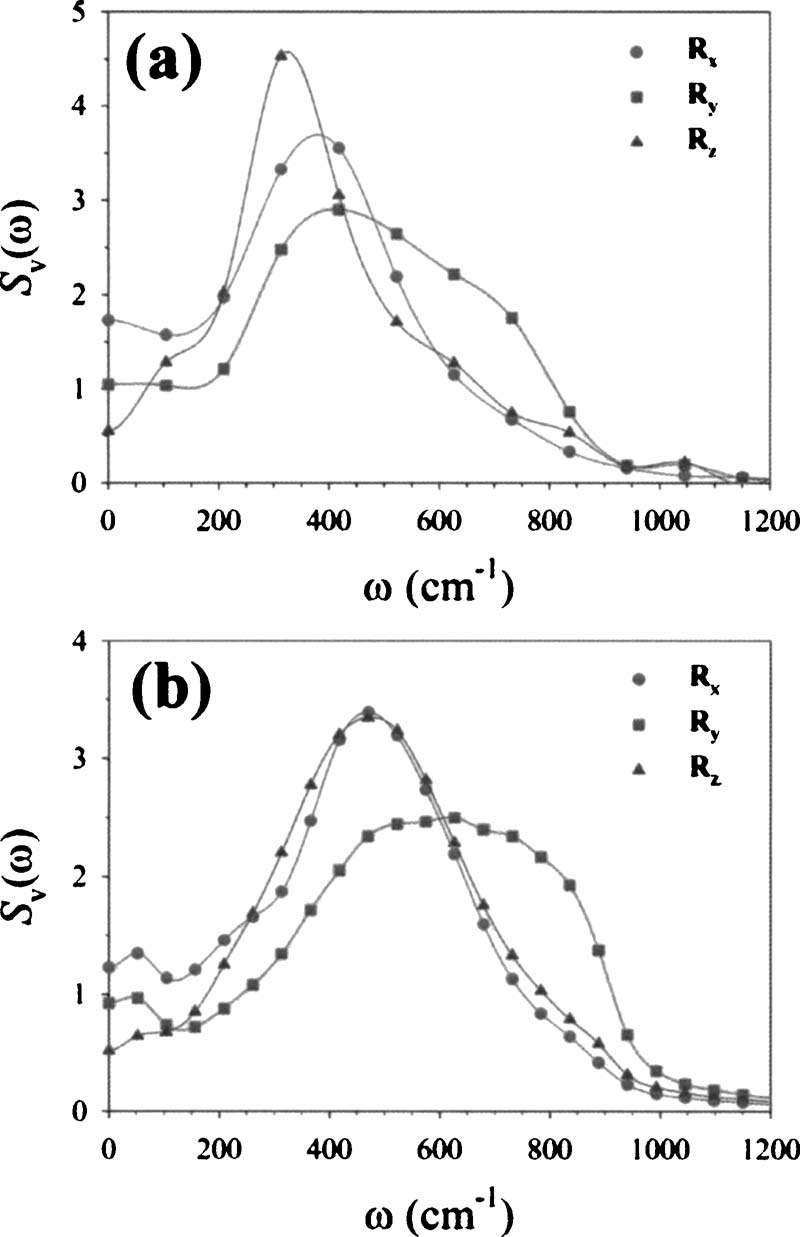}
\caption{Librational spectra of hydrating, (top, $r\e{e-O} \leq 3.7$~\AA), and `bulk' (bottom, $r\e{e-O} > 3.7$~\AA) water molecules. See appendix for computational details. Curves are obtained by a cubic spline interpolation.}
\label{fig:figure3}
\end{figure}

The radial distribution function between the electron position and surrounding water molecules reveals a hydration shell comprising approximately 4 water molecules~\cite{borgis02}. This hydration shell is defined according to the first minimum (at approximately 3.7~\AA~\cite{boutin03a,borgis02}) in the e$^{-}$-water RDF. To further investigate the origin of short-time electron dynamics we have evaluated the librational frequencies of both the hydrating and `bulk' water molecules. Figure~\ref{fig:figure3} shows the librational frequencies of hydrating ($r\e{e-O} \leq 3.7$~\AA), and `bulk' ($r\e{e-O} > 3.7$~\AA) water molecules at 298~K, computed according to the methodology of Bopp~\cite{bopp86} (see appendix). The peak frequencies, extracted from the interpolated data, are given in Table~\ref{tab:table1}. The dynamical behaviour of water molecules characterised as `bulk' is not greatly perturbed by the excess electron. For bulk water the restricted rotations (or rocking motio
 ns) that form the librational motions are manifested in vibrational spectra by the two bands, $L_1$ and $L_2$. Estimates of peak frequencies from IR and Raman spectra vary; with $L_1$ located around 380~cm$^{-1}$ and $L_2$ around 670~cm$^{-1}$ at ambient temperature~\cite{walrafen67,zelsmann95}. $R_{x}$ and $R_{z}$ constitute the lower frequency $L_{1}$ band and $R_{y}$ corresponds to the broader $L_2$ band.

The significant red-shift in the librational frequencies from `bulk' (Figure~\ref{fig:figure3}b) to hydrating water (Figure~\ref{fig:figure3}a) indicates that the electron perturbs the librational dynamics of the hydrating water molecules. This red-shift has been observed in both experiments~\cite{tauber03} and simulations~\cite{shkrob07}. A blue-shift in librational frequencies has been observed for hydrated ion systems~\cite{bopp88,rode02,rode97}, where the magnitude and direction of the frequency shift indicate ion-water interactions greater than water-water interactions. In this instance the weaker electron-water interaction red-shifts the librational frequencies.

On comparing the computed librational frequencies with the electron spectra in Figure~\ref{fig:figure2} the lower frequency peak (at 376~cm$^{-1}$) can be associated with hydrating water molecules (mainly $R_{x}$ and $R_{z}$ librations) whilst the higher frequency peak (at 653~cm$^{-1}$) can be attributed to non-hydrating water molecules (mainly the $R_{y}$ libration). These results confirm the long held notion, from experimental~\cite{bartels95,barbara98b,wiersma99,wiersma98} and simulation work~\cite{rossky86}, that the electron dynamics is dominated by water molecule librational motions. Furthermore, the direct correspondence between the librational and electron frequencies shows the electron to be highly responsive to individual water molecule motions. As such electron diffusion need not necessarily require collective water dynamics. 

Whilst the hydrated electron is a well-localised entity the fast oscillatory dynamics of the electron position suggest a highly fluxional local potential energy surface. In accordance with the conventional structural picture of electron hydration~\cite{kevan81} this potential energy surface is expected to be characterised by local minima near the oriented dangling H atoms of hydrating water molecules~\cite{borgis02}. The short-time dynamics observed in the particle coordinate (Figure~\ref{fig:figure2}) may then be understood as oscillatory movements between dangling H atoms. Electron motion being prompted by the fluctuating electrostatic field of the solvent librations. These rapid displacements between H atoms would be characterised by a free energy barrier on the order of $k\e{B}T$. The barrier itself being a function of the solvent coordinate (both the position and orientation of surrounding water molecules).

\begin{figure}[htb]
\includegraphics[width=8cm]{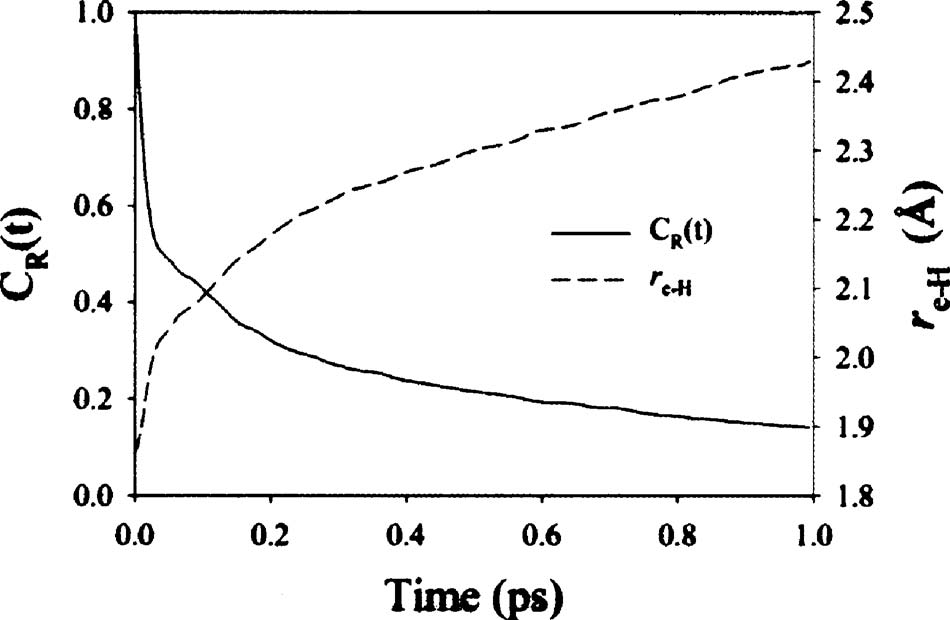}
\caption{Residence time correlation function, $C_{R}(t)$, computed at 298~K. Also shown is $r\e{e-H}$, the average distance between electron and H atom during the electron displacement.}
\label{fig:figure4}
\end{figure}

Computer simulations of Park et al. have demonstrated the importance of dangling hydrogen atoms in hydrated electron dynamics~\cite{park01}. They describe a process in which the electron, driven by dangling H atoms, responds to thermal fluctuations in the H-bond network. The destruction and formation of dangling H atoms in the surrounding H-bond network pushes and pulls the electron between water molecules. We assume that the diffusive process consists of a sequence of displacements between dangling H atoms. As such, the diffusion may be described by the exchange process identified in equation~\ref{equation1}. To compute the decay of $[A_{1},e^{-}\e{aq}]$ in equation~\ref{equation1} we define the following simple residence-time correlation function,
\begin{equation}
\label{equation5}
C_{R}(t) = \frac{\left\langle P(0) P(t) \right\rangle}{\left\langle P(0) P(0) \right\rangle}.
\end{equation}
The operator $P(t)$ is 1 if the electron, initially attached to a given H atom at $t=0$, is still attached at a later time $t$. The electron is defined as `attached' to the closest H atom. $C_{R}(t)$ is thus the residence time of the electron on a given H atom. The decay of $C_{R}(t)$ is thus the rate at which the electron moves between dangling H atoms. $C_{R}(t)$ is shown in Figure~\ref{fig:figure4} for 298~K. The initial fast decay of $C_{R}(t)$ ($t<0.05$~ps) reflects the fast oscillatory short-time dynamics. At longer times ($t>0.2$~ps) the short-time dynamics averages out revealing the slower long-time diffusive dynamics. Also shown in Figure~\ref{fig:figure4} is $r\e{e-H}$, the average distance between the electron and the H atom to which it was initially attached. From Figure~\ref{fig:figure4} we may estimate, by visual inspection, that electron displacements between dangling H atoms are within the range 0.7--1.0~\AA.

\begin{figure}[htb]
\includegraphics[width=8cm]{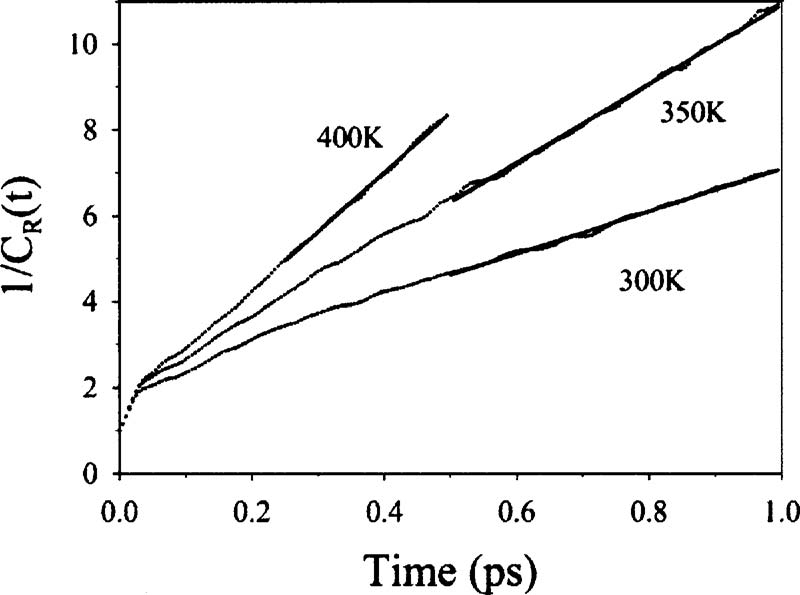}
\caption{Plot of 1/$C_{R}(t)$ against time at 298~K, 350~K and 400~K. Straight lines are linear fits to the long-time portion of 1/$C_{R}(t)$.}
\label{fig:figure5}
\end{figure}

The long-time proportionality of 1/$C_{R}(t)$ with time (Figure~\ref{fig:figure5}) confirms the second-order rate law typical of self-exchange processes. The gradient at long times gives the phenomenological rate constant, $k\e{exc}$, associated with the exchange process. The values of $k\e{exc}$ are listed in Table~\ref{tab:table2}. Noticeably, $k\e{exc}$ appears to be faster than the 1~ps timescale often associated with hydrogen-bond rearrangements in bulk water~\cite{fecko03,laenen98,luzar96}.

\begin{figure}[htb]
\includegraphics[width=8cm]{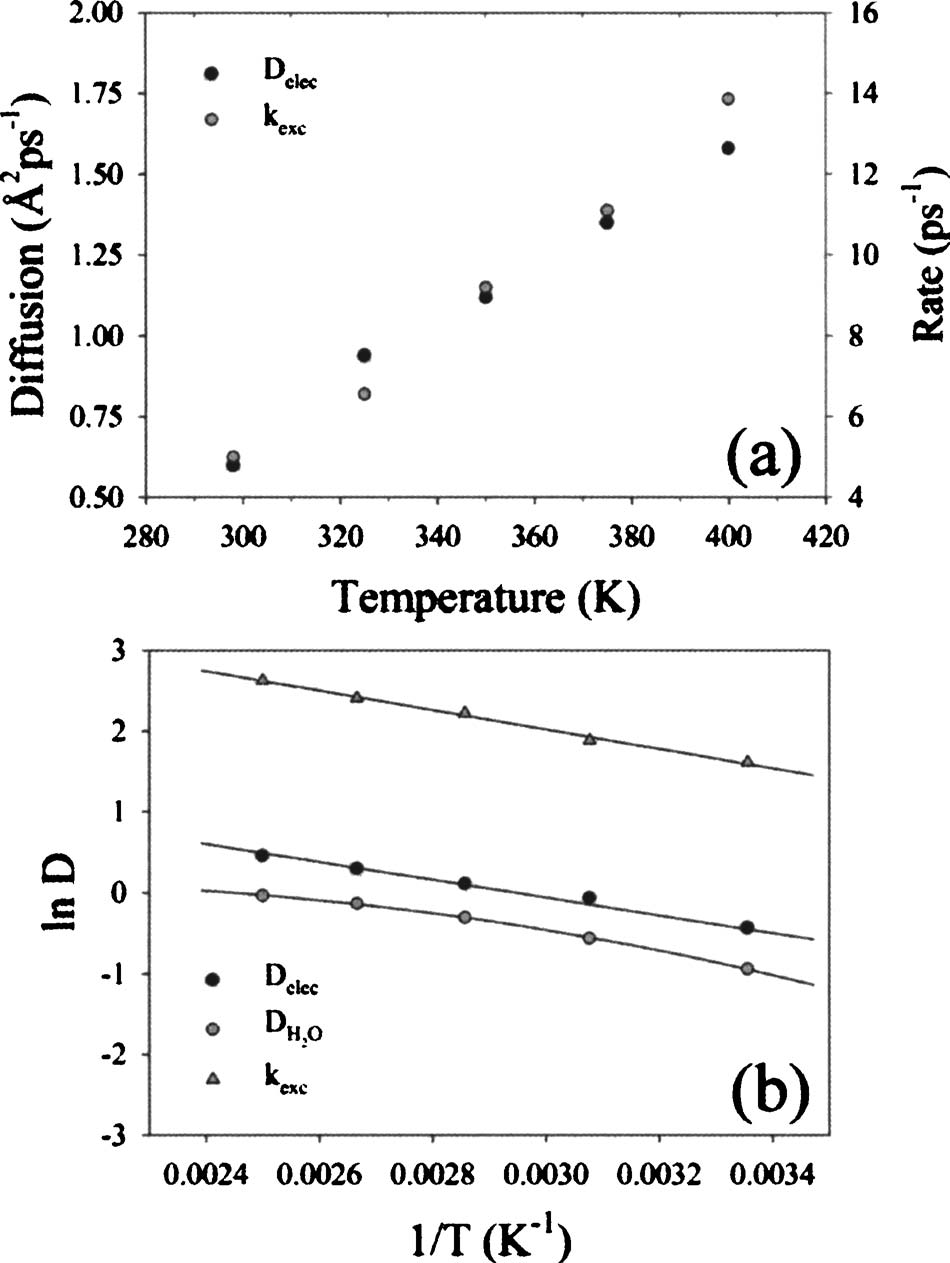}
\caption{Top, a plot of the electron diffusion coefficient, $D\e{elec}$, and the exchange rate constant, $k\e{exc}$, with temperature. Bottom, Arrhenius plots for $D\e{elec}$, $k\e{exc}$, and the bulk diffusion of SPC water, $D\e{H$_2$O}$.}
\label{fig:figure6}
\end{figure}

The calculated electron diffusion, $D\e{elec}$, and the exchange rate constant, $k\e{exc}$, are plotted in Figure~\ref{fig:figure6}a. Figure~\ref{fig:figure6}b shows that our model reproduces the Arrhenius-like electron diffusion. The Arrhenius behaviour of $D\e{elec}$, and $k\e{exc}$ constrasts strongly with the non-Arrhenius behaviour of the water self-diffusion. Clearly, neither the exchange process nor the electron diffusion are driven by the water diffusion. The (constant volume) activation energies for the electron diffusion and the charge transfer process are $8.9 \pm 0.7$~kJ mol$^{-1}$ and $10.0 \pm 0.4$~kJ mol$^{-1}$ respectively. The close agreement strongly suggests that the electron diffusion is well described by the exchange process in equation~\ref{equation1}. Experimental estimates of the activation energy are much larger at around 20.0~kJ mol$^{-1}$~\cite{bartels95}. 

\begin{figure}[htb]
\includegraphics[width=8cm]{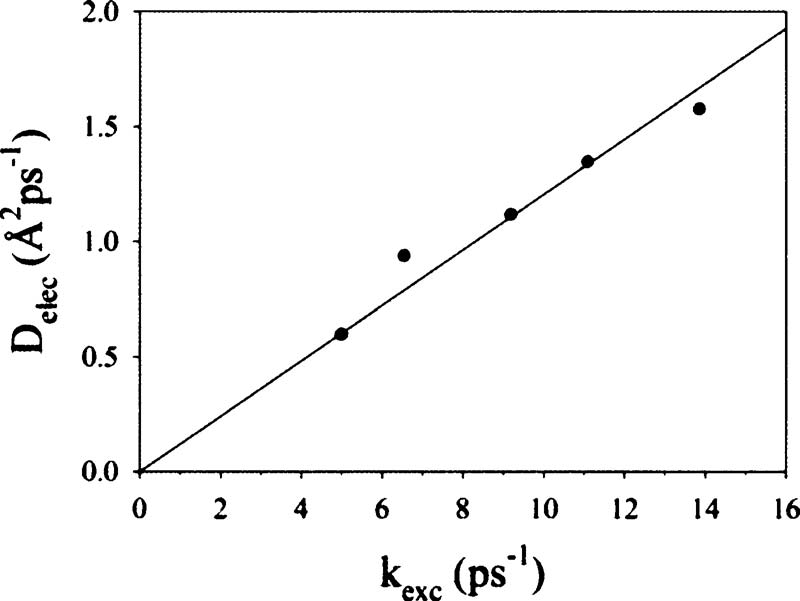}
\caption{A plot of $D\e{elec}$ against $k\e{exc}$. Also shown is the line of best fit through the origin.}
\label{fig:figure7}
\end{figure}

As a further test we assume proportionality between $k\e{exc}$ and $D\e{elec}$. We consider the Brownian behaviour of the electron as a simple random walk in which the step frequency is given by $k\e{exc}$,
\begin{equation}
\label{equation6}
D\e{calc} = \frac{\epsilon^{2} k\e{exc}}{6}.
\end{equation}
The parameter $\epsilon$ is the electron displacement accompanying the exchange process and is extracted from a linear fit of $k\e{exc}$ against $D\e{elec}$ (Figure~\ref{fig:figure7}). The displacement, $\epsilon$, is found to be 0.85~\AA{} and is in good agreement with our earlier estimates from $r\e{e-H}$ (see Figure~\ref{fig:figure4}). The good agreement between $D\e{calc}$ and $D\e{elec}$ (see Table 2) confirms that $k\e{exc}$ is indeed the phenomenological rate constant describing the hydrated electron diffusion.

In agreement with earlier works on electron diffusion~\cite{bartels92,murphrey92,park01} these results confirm the central importance of both the fast reorientational solvent motions and the instantaneous electron response. They also demonstrate that electron diffusion occurs rapidly at ambient temperatures purely under the influence of thermal fluctuations, and as such tunnelling processes are unnecessary to explain its high mobility~\cite{rossky89}. The model presented here reproduces the Arrhenius-like behaviour of the diffusion but underestimates the activation energy. The experimental Raman spectra of the hydrated electron is also qualitatively reproduced. A quantitative description of the electron dynamics requires a more accurate description of the fast solvent dynamics, including both intra-molecular degrees of freedom and polarisability. Comparison of computed and experimental spectra provides a clear test of the strength of the model~\cite{shkrob07,tauber03}.

The clear importance of librational dynamics on the electron diffusion raises an apparent paradox. The fast diffusion of the electron is enabled by the librational dynamics of the solvent. As such it may be expected that the increase in diffusion with temperature is due to an increase in the frequency of librational motions. However, librational frequencies decrease with temperature~\cite{zelsmann95}. The increase in diffusion appears to result from greater librational amplitudes and a more disordered hydration layer. Both may contribute to the fluxional character of the free energy barrier associated with electron displacements. Crucially, the respective back reaction of equation~\ref{equation1} is expected to diminish at higher temperatures as orientational configurations within the coordination sphere change much more quickly.

\section{Conclusion}
We have employed mixed quantum-classical molecular dynamics simulations to investigate the short and long-time dynamics of the hydrated electron within the temperature range 298-400~K. The short-time dynamics is characterised by oscillatory motions between dangling H atoms. This is a thermally activated process promoted by the librational dynamics of surrounding water molecules. Displacements between dangling H atoms constitute the fundamental step in the electron diffusion. We have demonstrated that the associated mechanism and kinetics can be understood as a transfer diffusion process in which the electron is exchanged between neighboring water molecules. The Brownian diffusion may then be modelled as a simple random walk in which the step frequency is given by the rate constant for the exchange process. The results presented here provide a consistent picture of both the short and long-time dynamics of the hydrated electron. They facilitate further investigations into the a
 pplicability of traditional ion diffusion models, water reorientation in the presence of anions\cite{Laage_anions}, and the seemingly anomalous hydrated electron diffusion at low-temperatures~\cite{bartels95} and within confined systems~\cite{boutin06a}.

\section{Acknowledgements}
The authors thank Bernard L\'{e}vy, Damien Laage and Rodolphe Vuilleumier for fruitful discussions.

\section{Appendix}
The three librational motions are labelled $R_{x}$, $R_{y}$, and $R_{z}$. The librations are approximated as vibrational modes corresponding to the instantaneous rotations of the water molecules about their three principal axes. Using the methodology of Bopp~\cite{bopp86}, we define two sets of unit vectors to describe the vibrational coordinates from which the modes are constructed. The first set of unit vectors, $\mathbf{\hat{v}_{1}}$ and $\mathbf{\hat{v}_2}$, assigned to H atoms 1 and 2 respectively, are perpendicular to the O--H bonds, within the plane of the molecule, pointing outwards. The second set of vectors, $\mathbf{\hat{p}_1}$ and $\mathbf{\hat{p}_2}$, are defined perpendicular to the plane of the molecule. The projection of the H atom velocities along the unit vectors $\mathbf{\hat{v}}$ and $\mathbf{\hat{p}}$ are labelled $V_1$, $V_2$, $P_1$, and $P_2$. We construct the normal coordinates $R_x$, $R_y$ and $R_z$ for a water molecule lying within the $yz$-plane wit
 h its $C_2$ axis aligned to the $z$-axis,
\begin{eqnarray}
\label{equation10}
R_x = V_1 - V_2 & \nonumber\\
R_y = P_1 + P_2 & \\
R_z = P_1 - P_2 &. \nonumber
\end{eqnarray}
$R_x$ decribes rotation about the $x$ axis; it is the in-plane swinging motion of the water molecule, often termed `rocking'. $R_y$, commonly referred to as the `wagging' motion, is a rotation about the $y$-axis during which the H atoms swing backwards and forwards across the molecular plane. Finally, $R_z$ describes the twisting of the water molecule about the symmetry axis. The spectral densities derived from the autocorrelation of the above normal coordinates provide the respective librational spectra. For further details readers are referred to the relevant literature~\cite{bopp86,bopp88}.

\newpage

\begin{table}[htb]
	\centering
		\begin{tabular}{c|c|c|c}
	\hline	
       & Hydrating (cm$^{-1}$) & Bulk (cm$^{-1}$) & Bulk Experiment (cm$^{-1}$) \\    
	\hline
	$R_x$  &  379  & 462 & 380 \\
	$R_y$  &  411  & 656 & 670 \\
	$R_z$  &  325  & 450 & 380 \\
  \hline
		\end{tabular}
	\caption{Computed librational frequencies for hydrating ($r\e{e-O} \leq 3.7$~\AA) and `bulk' ($r\e{e-O} > 3.7$~\AA) water molecules. Experimental values for bulk water are approximate~\cite{zelsmann95}. Peak frequencies are extracted from a cubic spline interpolation of the data set.}
	\label{tab:table1}
\end{table}

\begin{table}[htb]
	\centering
		\begin{tabular}{c|c|c|c|c|c}
	\hline
  Temperature & $D\e{H$_2$O}$ & $D\e{elec}$ & $k\e{exc}$ & $\epsilon$ & $D\e{calc}$ \\ 
  (K) & (\AA$^{2}$ps$^{-1}$) & (\AA$^{2}$ps$^{-1}$) & (ps$^{-1}$)  & (\AA) & (\AA$^{2}$ps$^{-1}$)  \\ \hline
	298  &  0.39 & 0.60  & 5.00  & 0.85 & 0.60 \\
	325  &  0.57 & 0.94  & 6.55  & 0.85 & 0.79 \\
	350  &  0.74 & 1.12  & 9.19  & 0.85 & 1.11 \\
	375  &  0.87 & 1.35  & 11.09 & 0.85 & 1.34 \\
  400  &  0.97 & 1.58  & 13.85 & 0.85 & 1.67 \\
  \hline
		\end{tabular}
	\caption{The self-diffusion coefficient of SPC water, $D\e{H\e{2}O}$, the electron diffusion coefficient, $D\e{elec}$, the electron exchange rate constant, $k\e{exc}$, the estimated random walk step distance, $\epsilon$, and the random walk model diffusion, $D\e{calc}$. As $C\e{R}(t)$ is dimensionless, the second order rate constant, $k\e{exc}$, is given as ps$^{-1}$.}
	\label{tab:table2}
\end{table}

\newpage

\end{document}